\documentclass[letterpaper, 10 pt, conference]{ieeeconf}  

\IEEEoverridecommandlockouts                              
\usepackage{balance}
\usepackage{cite}
\usepackage{balance}
\usepackage{cite}
\usepackage{amsmath,amssymb,amsfonts}
\usepackage{algorithmic}
\usepackage{graphicx}
\usepackage{textcomp}
\usepackage{todonotes}
\usepackage{cleveref}
\usepackage{url}

\title{\LARGE \bf Using AI/ML to gain situational understanding from passive network observations
\author{Dinesh Verma$^{1}$ and Seraphin Calo$^{2}$}
\thanks{$^{1}$D. Verma is with IBM T. J. Watson Research Center, Yorktown Heights, NY 10598, USA {\tt\small dverma at us.ibm.com}}
\thanks{$^{2}$S. Calo is with IBM T. J. Watson Research Center, Yorktown Heights, NY 10598, USA {\tt\small scalo at us.ibm.com}}
}
\date{2017\\ December}

\begin{document}

\maketitle
\thispagestyle{empty}
\pagestyle{empty}

\label{sec:abstr}

\begin{abstract}
The data available in the network traffic from any Government building contains a significant amount of information. An analysis of the traffic can yield insights and situational understanding about what is happening in the building. However, the use of traditional network packet inspection, either deep or shallow, is useful for only a limited understanding of the environment, with applicability limited to some aspects of network and security management. If we use AI/ML based techniques to understand the network traffic, we can gain significant insights which increase our situational awareness of what is happening in the environment. 

At IBM, we have created a system which uses a combination of network domain knowledge and machine learning techniques to convert network traffic into actionable insights about the on premise environment. These insights include characterization of the communicating devices, discovering unauthorized devices that may violate policy requirements, identifying hidden components and vulnerability points, detecting leakage of sensitive information, and identifying the presence of people and devices. 

In this paper, we will describe the overall design of this system, the major use-cases that have been identified for it, and the lessons learnt when deploying this system for some of those use-cases. 
\end{abstract}
\section{Introduction}
\label{sec:intro}


Almost all buildings in any government, military or commercial enterprise today operate using a network which communicates using the Internet Protocol~\cite{rfc791}. There is significant information available in the network packets that are travelling back and forth between the occupants of the building, and to the different machines outside the building. 

This network traffic has been mined for information, but the primary applications for which it has been used is only to be found in network security ~\cite{li2013survey} for use-cases such as intrusion detection and intrusion prevention. The other primary use-case for analyzing network traffic has been in creating network traffic analysis models, which can find applications in network planning and deployment.  

However, network packet inspection has many uses which go beyond the scope of network security analysis network planning. The analysis of network packets can provide useful 'situational awareness' of what may be happening within the network, identifying people and devices that are present in the environment, vulnerabilities in the network infrastructure, behavior of devices, and several other uses. Obtaining much of this insight from the network packets requires combining pre-existing knowledge with the content being carried within the network, using a mixture of machine learning algorithms along with some domain knowledge of networks, and a flexible infrastructure that can support a variety of use-cases. 

In this paper, we discuss these broad set of use-cases which can be supported using network packet analysis, and describe a system which we have built to implement these use-cases. A key aspect of successful determination of the situation in these cases requires a a combination of network domain knowledge and an application of AI technologies. We will discuss both aspects in this paper. 
\section{Use-Cases}
\label{sec:use-cases}
\begin{figure}[thpb]
      \centering
      \includegraphics[scale=1.0]{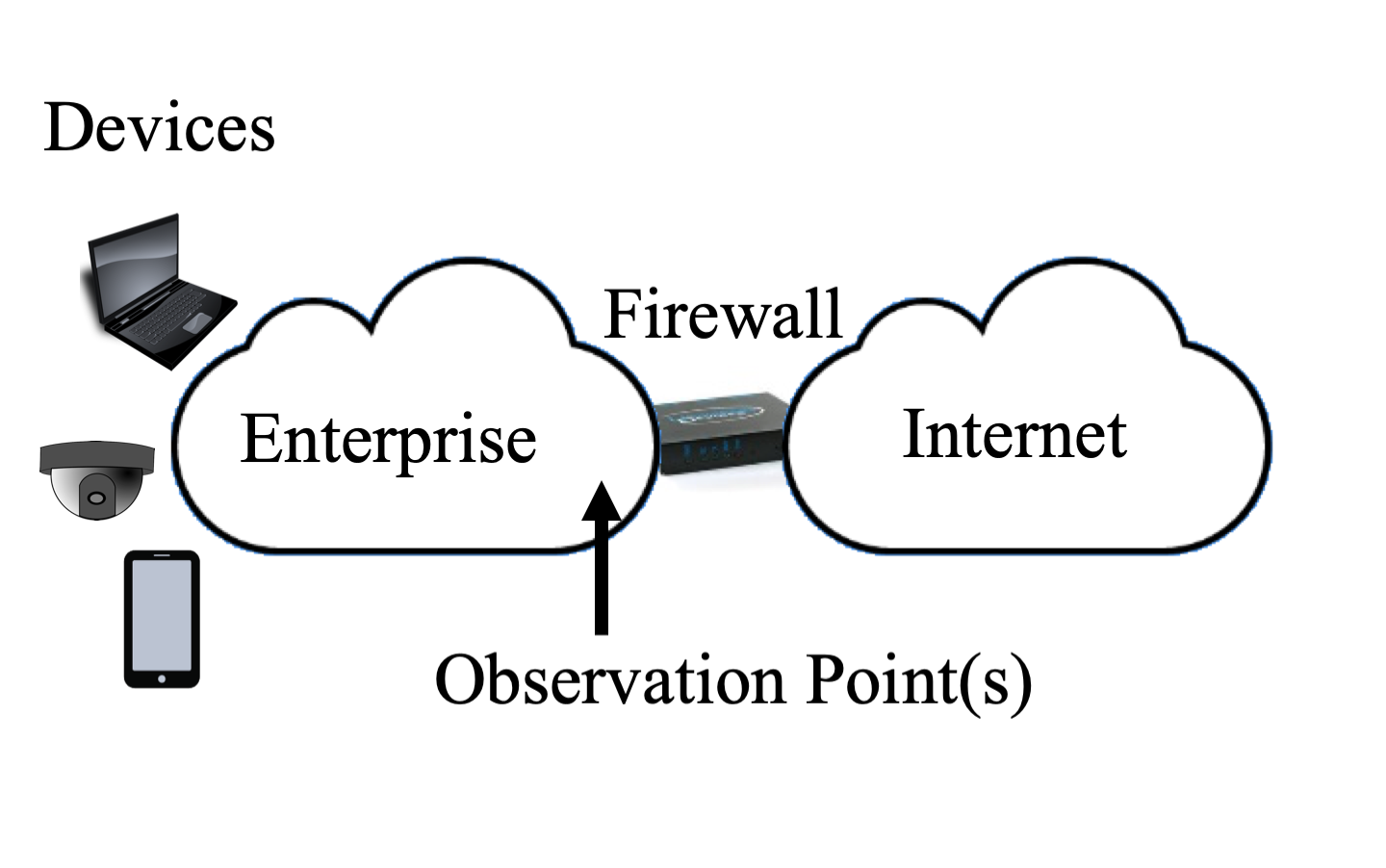}
      \caption{The environment assumed for Cyber Physical Systems. Cyber phyiscal systems are connected and made accessible over the Internet. The most common usage will be a manager or legitimate user issuing control commands to the IoT device. }
      \label{fig:fig1}
\end{figure}

The typical deployment scenario we have is shown in Figure ~\ref{fig:fig1}. The network packet collection system is installed in a building (or other premises) which is used to monitor packets just before they exit the firewall. The packet monitoring system can also collect information from other points in the building, if needed. The system may further incorporate data available within the building or outside it to provide auxiliary information, such as any information about registration of devices with users, or information available from some global location. 

Because the observation point is within the firewall, we assume that the devices in the network can be identified by their IP addresses. The IP address assignment of any device could change, but the change can usually be detected if the system also has information from other sources such as the DHCP servers ~\cite{droms1999automated}  deployed within the environment. Depending on the deployed environment and the location and number of observation points, there may be alternate ways of identifying the device from its IP address. 

Packet inspection tools such as Zeek~\cite{zekeref}, snort~\cite{roesch1999snort} or wireshark~\cite{orebaugh2006wireshark} can provide an initial examination of the contents of the packets, but their functions can be augmented with AI and machine learning capabilities. With this augmentation the observation of network packets can lead to the following set of information. 

\vspace{1mm}

\textbf{Discovery of devices:} The number, type, and attributes of devices that are connected to the network and communicating actively on it can be detected, identified and characterized. These include devices which do not have a management agent installed on them, and do not respond to queries initiated by network management protocols such as SNMP, or system management protocols such as ~\cite{stallings1999snmp} WEBM ~\cite{thompson1998web}. The only devices that can not be identified by this mechanism are those that never communicate on the network. 

The discovery of devices is important for a variety of purposes. For one, it lets the administrators identify any unauthorized devices that are present in the environment. Also, in the case of network audits, it provides a report of all the devices present. Some enterprises conduct an audit of all devices on a periodic basis, and the discovery of devices is an important component of that audit. In some cases, the number of devices present in an environment of a particular type may be required to check on the number of licenses needed for their use. In other cases, when the operation and management of an environment needs to be turned over to an outsourcing company, the discovery can provide a better estimate of the effort and cost required for the outsourcing. 

\vspace{1mm}

\textbf{Detecting Policy Violations:} When devices are identified, they can be identified along with their attributes such as their manufacturer, model, operating system, firmware version, etc. The protocols used by the device to communicate on the network can also be identified. Inspection of network packets can be used to validate that devices are not violating any policies that are specified on the use of the network. Typical policies may include the requirement that all communicating devices be registered in a database, or that all devices use secure communication. Other types of policies may prevent accessing some class of websites, or sites within some specific geography. These violations of policies can be identified by means of packet observation, and checking them against the list of registered devices, or checking the communication protocols they are using. If a class of devices is not allowed on the network, e.g., an enterprise may want to disallow recording devices like Alexa within its buildings, any policy violation can be detected if such a device is observed communicating on the network.  

\vspace{1mm}

\textbf{Discovering network topology:} Network topology at the IP layer (Layer 3) of connectivity can usually be determined fairly easily, e.g. by means of probing network devices using snmp ~\cite{pandey2009ip}. However, active probing by a system can generate loads on the network, interfere with the operations of the network, and may not work in some environments where multiple tiers of packet filtering firewalls are installed. However, discovering the physical (Layer 2) network topology, i.e., identifying things using their physical MAC addresses and how they are interconnected is much more difficult. Passive network management schemes which can capture and analyze MAC addresses and header fields can be used to construct physical (Layer 2) network topology. 

Even in the construction of network topology at the IP layer (Layer 3), passive network measurements can be useful. While such approaches have been used to understand and construct the topology of the Internet ~\cite{eriksson2008network, donnet2007internet}, they have not been typically used to understand the structure of an enterprise network. However, combining the topology inferred from observation of network traces at different points within the network can help in constructing the topology of an enterprise network. Such passive network discovery tools can be useful for understanding the connectivity between different devices, routers and end-points within an enterprise.   

\vspace{1mm}

\textbf{Understanding network resiliency:} One of the advantages of understanding network topology is that it can be analyzed to understand components that may be vulnerable, or those whose performance degradation can cause a significant impact on the performance of other systems. In order to understand and identify such components, we need to determine not just the topology, but also understand the traffic characteristics among the different servers and machines within the network. Additionally the inter-dependencies among the protocols used in the traffic needs to be identified as well. 

\vspace{1mm}

One of the challenges in understanding network resiliency is the task of identifying hidden components. A hidden component is an element in the infrastructure which other elements may be dependent on, but which may be overlooked in the task of system and network planning and upgrade. As an example, upgrading servers supporting an application while ignoring to update the capacity of the systems involved in backup of data for that application, can cause performance degradation that may go undetected. Similarly, installing a spam checker for email which performs inverse domain name lookup to validate email sources can slow down overall application performance if corresponding performance upgrades to domain name services are not made. Such hidden components, which can impact network resiliency or network performance, can be identified by passive network observation. 

\vspace{1mm}

\textbf{Understanding Device Behavior:} Many modern devices can operate in different modes. As an example, tablets may be used for surfing the web by users in an interactive manner, and the same type of tablet may be embedded as a controller within a network printer, or be used as a sensor capturing video or sounds in some space. Likewise, smart televisions used in conference rooms can serve as display devices for presentations or streaming video, or be used as browsers to display information from the internet. An examination of the network packets, and identifying the network communication behavior of the devices can help identify which behavior is being exhibited by a device in any period of time. This examination of behavior helps both identify the type of device and provide greater insights into the role of different devices within the enterprise. 

\vspace{1mm}

\textbf{Understanding Presence:} In many buildings, it is useful to keep track of how they are being used, how many occupants are present in the building at any given time, and how the occupancy of the building changes over time. Various approaches to estimate occupancy using visual and vibration sensors ~\cite{shih2014robust,pan2014boes} as well as chair occupancy sensors ~\cite{labeodan2016experimental} have been proposed. However, in addition to the intrusive nature of such sensors, the cost of deploying and managing them is very high. Passive network packet inspection to detect the presence of people provides an alternative and low-cost approach to determine how many occupants are within a building. The premise behind this estimation is that all occupants would be communicating over the network using one or more devices, and associating the identify of the devices with that of the user can provide an accurate count of the occupants within the building. 

These are but a few of the use-cases that can be supported using passive network observations. In the next section, we look at the architecture of a system that can provide a common way to support these use-cases. 
%
\section{Architecture }
\label{sec:arch}
It was recognized that there was no one unique method of analysis that could be used to determine the various factors that comprise situational awareness. The system was thus structured to be able to employ multiple types of analytics to determine the relevant attributes of particular devices communicating on the network and the relationships between them. The ability to include both algorithmic and AI based components gives the system a wider scope and greater effectiveness.

\begin{figure} [thpb]
      \centering
      \includegraphics[scale=.275]{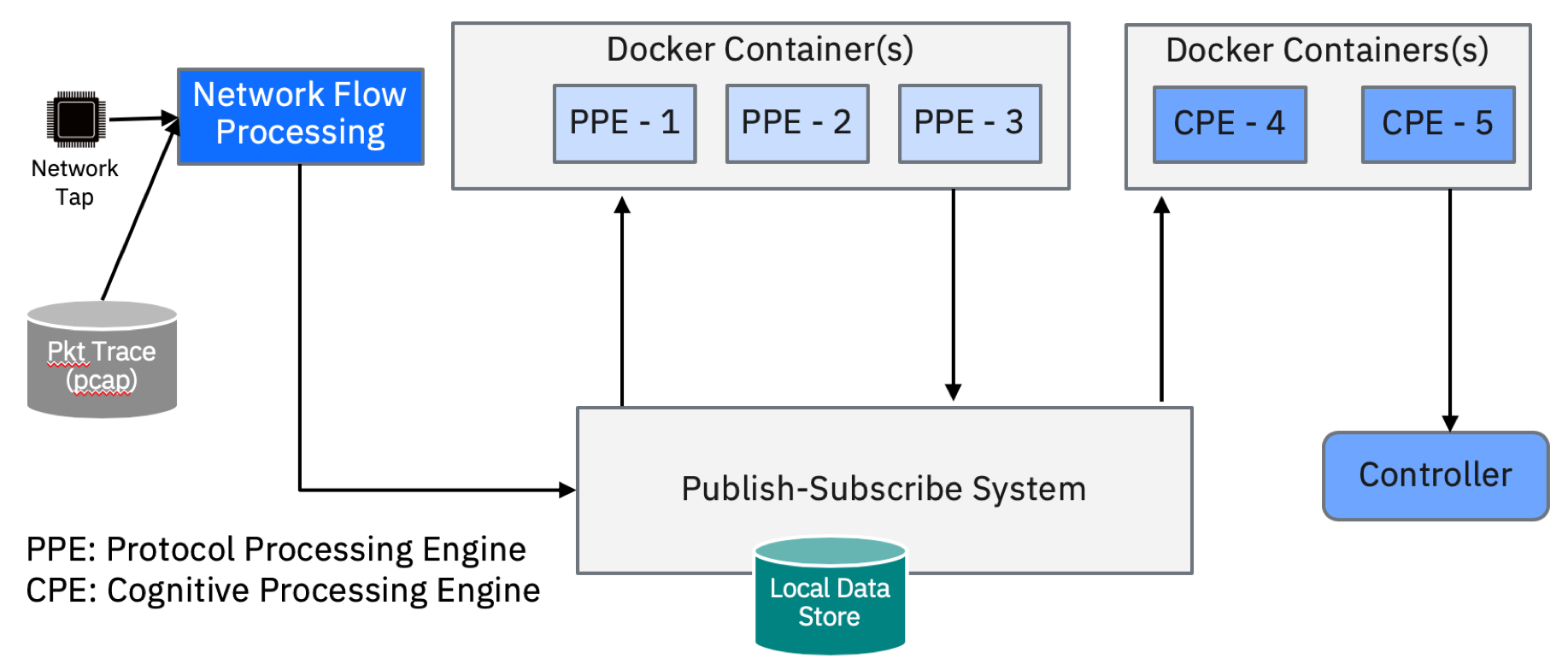}
      \caption{Analyses are performed by chains of algorithmic and machine learning components}
      \label{fig:chain}
\end{figure}

\begin{figure} [thpb]
      \centering
      \includegraphics[scale=.35]{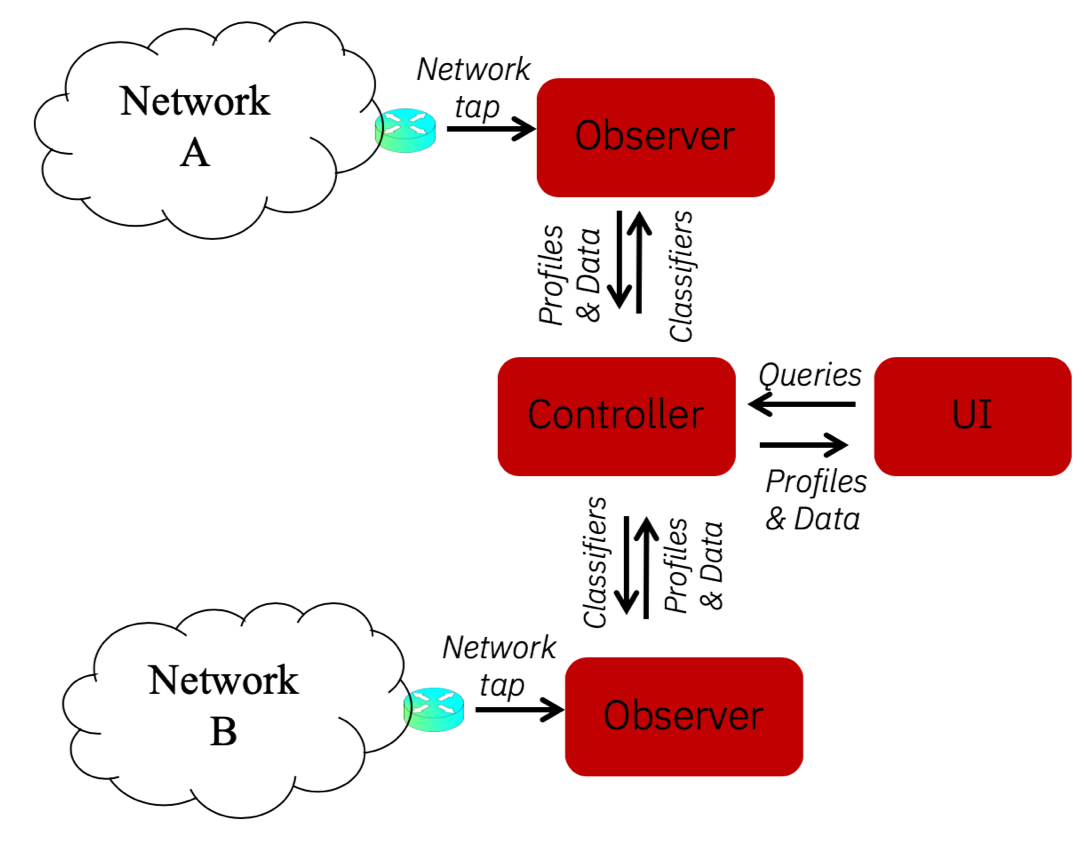}
      \caption{The high level architecture}
      \label{fig:hla}
\end{figure}

The analysis components are structured in chains as in Figure~\ref{fig:chain} that can be either individually selected or run in parallel against the same data. The data input consists of packets either streaming directly from the network or previously captured in pcap files. The results obtained from each analysis component are captured in a file that ultimately contains the combined contributions of each chain. The information keeps on getting enriched as the data processing continues along the chain, and the component elements can use the results of all earlier analyses in their processing. The data objects containing the processed information are referred to as Profiles in Figure~\ref{fig:hla}.

The system architecture follows an edge computing paradigm wherein AI models are created and trained in the cloud, and then sent to an observation device at the edge that maintains the processing chains. The edge devices monitor network traffic and perform the analyses for which they have been configured.

The resulting system thus consists of two main classes of software elements as shown in Figure~\ref{fig:hla}: observers and controllers. Observers are the elements that are located at the edge devices and provide the system observation functionality. The observers run the chains of analysis modules, which are divided into two types, Protocol Processing Engines (PPEs) and Cognitive Processing Engines (CPEs). PPEs are based on analytics that examine and interpret the known characteristics of the protocols used between communicating elements of the system. CPEs embody machine learning techniques based upon behaviors reflected in previously captured training data. The results of all the analyses are sent to the controllers, where they are combined and stored. 

The composition of the outputs of the analysis chains can be done in a number of different ways. Results pertaining to the same attributes must be resolved. The collection of components can be viewed as an ensemble with all relevant classifiers contributing to the final decision on each attribute. This assumes that the classifiers are diverse and have competitive accuracies. Alternatively, the most accurate classifier for each attribute can be chosen to make the decision if it is known. In systems where the ground-truth can be ascertained, this information can be used to provide a scoring set for determining the accuracy of the different chains.

The controllers are located in the cloud and coordinate the activities of the observers. They can also perform analysis functions that may be too computationally expensive to be done at the observers. Typically, there would be one controller for many observers.

\section{Implementation}
\label{sec:impl}

The implementation of the system consists of open source components, customized protocol processing elements, and AI-enabled processing elements. Where possible, we have utilized open-source technologies to speed our development; and, for ease of deployment and operations, all components are instantiated in Docker Containers. 

\begin{figure} [thpb]
      \centering
      \includegraphics[scale=.34]{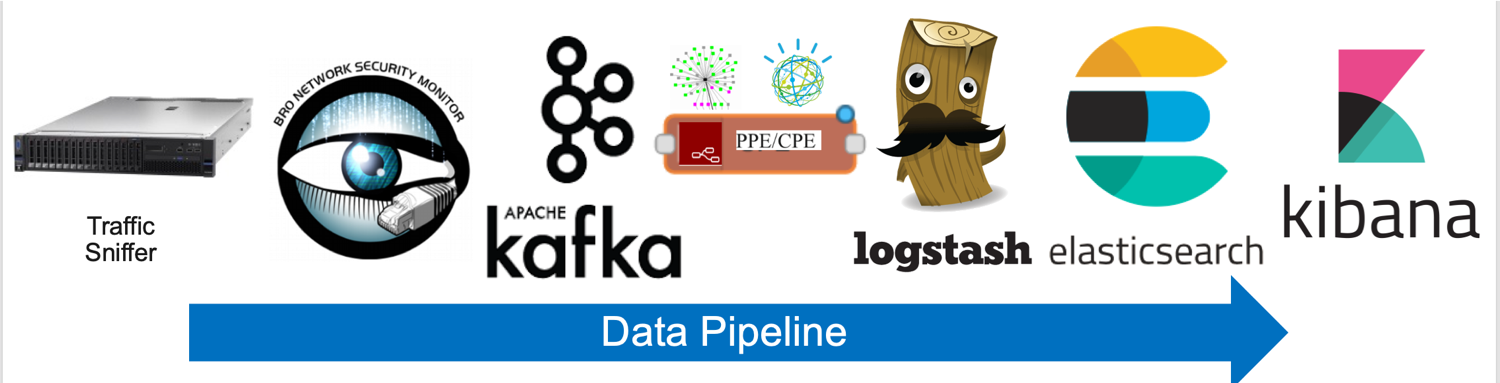}
      \caption{Data Pipeline}
      \label{fig:pipe}
\end{figure}

Analysts get insights concerning the behavior of the devices communicating over the network being observed by using the data pipeline shown in Figure~\ref{fig:pipe}. This pipeline is designed to support data at scale, and filters and reduces the amount of data as it proceeds through the multiple stages of capture, analysis, integration, and interpretation. This supports a view of the "big picture" while allowing details to be preserved as needed.

\vspace{1mm}

\textbf {Zeek}~\cite{zekeref} (formerly Bro). We use the Bro open-source software network analysis framework for the basic extraction of information from the network flows. It is a passive packet capturing system most often used for Intrusion Detection. We use Bro 2.5.5, configured to capture Connection, DHCP, DNS, HTTP, and SSL/TLS traffic information as well as x509 certificate data. Extensions were added to provide additional information, and the Kafka plugin was used to send information to the message bus. We have used this software to capture data from the IBM Yorktown building network which produces roughly 8 GB a day, to help stress the system and demonstrate its capabilities.

\vspace{1mm}

\textbf {Kafka}~\cite{kafkaref}. For a distributed streaming platform we use Apache Kafka. It supports multiple producers and consumers, similar to a message queue or enterprise message bus. Kafka stores a data stream for a configurable amount of time, e.g., a day or a week, and then deletes it. We do not use it to persist the data, but rather to get data from one application to another. This allows very flexible and configurable interconnections within the system.

\vspace{1mm}

\textbf{Database Resources}. Various repositories of information are utilized to provide additional knowledge assets for the analysis components. For example, one DB is a reverse-name lookup table of IP to DNS address mapping, another is a MAC address OUI lookup table identifying vendors of specific equipment, and a third is information on the ISP owner of a particular network. Such assets are typically invoked via remote procedure call by the PPEs and CPEs to meet their analysis needs.

\vspace{1mm}

\textbf {Protocol Processing Elements} (PPEs). These are custom applications that read from the Bro Kafka topics, transform the data in some way, and then write to a new topic on Kafka. They examine and interpret the known characteristics of the protocols used between communicating elements of the system. The protocol parsers focus on key events and extract the semantics behind the stream of bytes. They typically are used to examine protocol and address information from the packets flowing in the network; map protocol data from various sources into a common data abstraction; and, extract key data from the traffic flow that can be used in support of the analysis components.

\vspace{1mm}

\textbf{Cognitive Processing Elements} (CPEs). CPEs are pro- cessing elements that include AI capabilities. They are the heart of the system, and perform the analyses that provide the insights about the communicating devices and their behaviors. CPEs have been developed for: determining whether a device is an IoT device; mining data from DNS look-ups; invoking Internet services like WhoIs and IP2Location for identifying characteristics of source and destination addresses; inspecting HTTP user agent and URI strings for device details; analyzing TLS certificate information for encrypted traffic; and, interpreting time-series characteristics of the network data streams.

\vspace{1mm}

\textbf{Logstash}~\cite{elasticref}. Logstash is a data collection engine that can dynamically unify data from disparate sources and normalize the data that is sent to designated destinations wherein it can be parsed and processed as required. Logstash is an extract, transform and load (ETL) tool that converts between many formats. In our case, we transform from Kafka topics into the ElasticSearch database.

\vspace{1mm}

\textbf{ElasticSearch}~\cite{elasticref}. Elasticsearch is a powerful search and analytics engine. It provides a distributed, multitenant-capable full-text search engine with an HTTP web interface and schema-free JSON documents. ElasticSearch is where data is persisted, to be queried and analyzed by a user.

\vspace{1mm}

\textbf{Kibana}~\cite{elasticref}. Kibana is a data visualization plugin for Elasticsearch. It provides visualization capabilities on top of the content indexed on an Elasticsearch cluster. It has a Web interface for graphing, charting, and visualizing data, and allows composing visuals into dashboards that expose a high- level interface into the data. 

\vspace{1mm}

The system was designed to be extensible, leveraging a distributed edge/cloud architecture, and sophisticated analysis components connected together by a messaging bus to enable easy addition/removal/creation of new modules. The components can be packaged into a variety of hardware configurations, from a single network appliance to a cluster of machines thus making the system adaptable to different environments and uses.
The overall design was split into a Controller and multiple Observers (cloud and edge functions) to make it scalable. Heavy weight operations can be performed at the Controller, while the collection of Observers deal with high-volume traffic in real-time.

Analyses are performed using a combination of Domain Knowledge and Machine Learning. The Protocol Processing Engines leverage the structured nature of the communications information and learn by exploiting network domain knowledge. The Cognitive Processing Engines use the power of AI to understand patterns in the network traffic. Chains of CPEs and PPEs capture new insights from the analysis of different attributes of the communicating devices or from pursuing different perspectives in the analysis of the same attributes. Various queries can be formulated to explore relationships among the different devices in the system, and their results captured as part of the overall characterization of the network activity.

\section{Extraction of Knowledge}
\label{sec:knowledge}

In this section, we discuss how the various use-cases described in Section ~\ref{sec:use-cases} are attained using the architecture and implementation described in the other sections. In order to support the use-cases, the processing elements in the system need to extract different types of knowledge from various protocols, combine that with other existing pieces of knowledge, and eventually produce the desired output (namely a profile that contains all the fields required to address the use-case). 

The knowledge required for each of the use-cases can come from either external knowledge sources, or be captured from the network traffic.

\subsection{External Knowledge Sources}
There are several sources of knowledge which are available to a traffic analysis system as it analyzes the traffic passing through the network. These external sources of knowledge can be combined with the information carried within the network packets to address the various use-cases. These external sources of knowledge include: 

\vspace{1mm}

\textbf{Domain Ownership Information:} Different domains in the Internet are owned by different organizations. There are a variety of tools available which allow one to retrieve information about domain name ownership, such as whois~\cite{daigle2004whois} which allow one to retreive details about the ownership of a domain name, which includes the identity of the owning organization, the location of the machine if the domain name is that of an individual machine, and information about the technical and administrative contact personnel. In some cases, the information may be obscured, but for the case of most large manufacturers of devices, the whois record can provide the identity of the owning organization. 

\vspace{1mm}

\textbf{Geographic Location of Addresses:} The information contained in the network traffic only identifies the IP address of communicating end-points. However, a variety of techniques ~\cite{padman2001} 
exist to map those network addresses to geographic locations in the real world. While the accuracy of such techniques is far from perfect ~\cite{poese2011ip}, they do provide a crude estimate of where a specific address in the network may be located physically. 
\vspace{1mm}

\textbf{Signature Descriptions:} In many cases, the attributes of a device talking on the network can be identified by means of rules that map specific patterns seen in network traffic to information about the origin of the device. An example of such a signature mapping is the value of the 'User-Agent' field carried as a header within the HTTP protocol. While a large number of these user-agent fields exist, the type of device and the type of browser running on a client can be determined from the network traffic if a rule mapping this field is available during network traffic processing. Information mapping various value of this field to attributes of browsers and devices is available from some web-based sites ~\cite{ualistsite}. Similarly, the type of payload contained in communication can be identified using the specific markers in an audio or video file payload, or the set of typical suffixes used to describe a file. These are also available within various pages on the Internet, e.g., a set of common prefixes for audio is listed in Wikipedia ~\cite{wiki-audio}. These sources of information can be used to generate a set of rules that can identify the type of communication that is happening. 

\vspace{1mm}

\textbf{Work Stations and Device Registry:} In many enterprises, a registry of ownership information about devices is maintained, which would typically record the serial number of a workstation along with the name of the owner. This registry serves as a means of establishing a relationship between a specific device and a member of the enterprise organization. In some enterprises, a registry of static IP addresses and the machines to which they are assigned is present, which can allow one to understand the configuration of the network. 

\vspace{1mm}

\textbf{Network and Systems Management Databases:} In many large enterprises, it is common to have a system administrator site, along with or in conjunction with a network management site and a Network/Systems Operational Console. These management systems would typically use the management agents on different devices to collect information about the current topology of the network and systems, track issues that users may be having concerning performance of the applications, as well as a record of changes in the configuration or upgrades on different network software or related information. This information can be combined with network traffic to get additional insights.  

\vspace{1mm}

These external sources of knowledge are combined with the knowledge that is carried in the network traffic in order to address the specific needs of each use-case. 

\subsection{Information in Network Traffic}
There is a substantial amount of information that can be gleaned from the headers of various protocols in the network. A subset of that information is listed below for a selected subset of protocols.  

\vspace{1mm}

\textbf{DHCP:} The DHCP (Dynamic Host Configuration Protocol) is used by computers to get their IP addresses in an environment. They usually provide their identification information when retrieving this address, which could include a certificate, the MAC address of their network interface, or the identity of the user owning the machine. The records of the DHCP server assignment, or an examination of the packets travelling to and from the DHCP server provides information to associate a unique identity with an IP address. When the IP address is reassigned, the DCHP server would usually be able to tell who the new IP address belongs to, and provides the marking point which can separate two devices that happen to own the same IP address. 

\vspace{1mm}

\textbf{DNS:} The domain name server captures all requests that are made by a machine to translate a domain name to an IP address. In an era of content distribution networks and virtualized hosting, an examination of the DNS requests allows one to map an IP address to the domain name it is being used for. Furthermore, by combining it with the domain ownership information, the system can map an IP address to its location, or to the organization that owns it. Note that this mapping needs to be built up dynamically in the network since the existence of content distribution networks, wide area load balancing and virtual hosting means that the static information can not be used directly. 

The patterns of requests made to the domain name service by a machine also help to identify its attributes and behavior. IoT devices which are usually single-function will make calls mostly to domains owned by their manufacturer, and that information (coupled with domain name ownership) acts like a good marker for the manufacturer of a device. The specific type of domain that is accessed can reveal what type of device the machine is~\cite{le2019policy}. 

\vspace{1mm}

\textbf{IP:} The IP protocol encodes the source and destination addresses of the packets. When combined with the information available from the domain name services, and the geographic mapping, this provides information about where the communication is happening. Furthermore, clustering analysis on packets and destinations can identify anomalies, identify heavy congestion points, and identify points of vulnerabilities and criticality in the network. 

\vspace{1mm}

\textbf{TCP:} The transport protocol encodes the source and destination ports used by the application, and can identify the typical applications that are being used within the network. This coupled with the IP level statistics can provide additional insights into anomalies and critical components of the network. 

\vspace{1mm}

\textbf{TLS: } The TLS protocol aims primarily to provide privacy and data integrity between two or more communicating computer applications. Even though the payload in the TLS traffic is encrypted, there exist various ways to identify characteristics of devices that use TLS for communication. The exchange between the devices includes certificates provided by the websites and in some cases client-side certificates are carried. By identifying the issuer of these certificates, the identity of the manufacturer of a device can be inferred. Cipher-suites proposed for TLS communication also yield information about the nature of the originating device, e.g., what application stack is being used.

\vspace{1mm}

\textbf{HTTP:} HTTP is an application protocol that is the foundation of the World Wide Web. Hypertext documents include links to other documents which can be accessed via the HTTP protocol. The headers used in the protocol  allow identification of the attributes of the end-points, including its operating system, the type of browser used, and other attributes.  For example, the user agent string field in the HTTP header is specified by the software making an HTTP request to describe the capabilities of the client device. Similarly the target of each HTTP request is a resource which is defined by a Uniform Resource Identifier (URI).

\vspace{1mm}

Other protocols such as MODBUS, SCADA and BACNET that are used for communication in specialized building management systems also reveal information about their users and devices. 

The combination of network protocol knowledge and the external knowledge sources provides enough details to support all the use-cases described earlier.

\section{Conclusions}
\label{sec:conclusion}

We have presented an overview of a system that performs passive network traffic analysis using AI/ML. The combination of the techniques described along with network domain based packet examination can unlock the knowledge about the situations occurring in an enterprise or a facility, enabling a variety of use-cases ranging from discovery of devices to detecting presence of individuals and estimating occupancy of buildings. 

The system is structured to be very flexible and adaptable. By changing the analytics and the manners in which they are interconnected, it can be applied to a range of different problems. The system can thus be extended to support other use-cases that we have not mentioned in this paper. 

\bibliographystyle{IEEEtran}
\bibliography{references}
\balance

\end{document}